\documentclass[twocolumn,prd,
 reprint,
nofootinbib,
amsmath,amssymb,
aps
]{revtex4}

\usepackage{tikz}
\usetikzlibrary{shapes,fit}
\usepackage[applemac]{inputenc}  
\usepackage{graphicx} 
\usepackage{epsfig}
\usepackage{hyperref}
\usepackage{amssymb} 
\usepackage{float}
\usepackage[percent]{overpic}
\usepackage{changes}

\usepackage{amsmath}
\usepackage{mathrsfs}

\usepackage[english]{babel}
\usepackage{setspace}

\renewcommand{\vec}[1]{\boldsymbol{#1}}
\DeclareMathOperator{\tr}{Tr}
\DeclareMathOperator{\Rt}{\vec{R}^{\left( s\right)}_{tt'}}
\DeclareMathOperator{\R}{\vec{R}^{\left(s\right)}}

\begin{document}
\title{Dynamic system classifier}

\author{Daniel Pumpe\footnote{dpumpe@mpa-garching.mpg.de}$^{1, 2}$, Maksim Greiner$^{1, 2}$, Ewald M{\"u}ller$^{1,3}$ and Torsten A. En{\ss}lin$^{1,2}$}
\affiliation{$^{1}$ Max-Planck-Institut f\"ur Astrophysik, Karl-Schwarzschild-Str.~1, D-85748 Garching, Germany\\
$^{2}$ Ludwigs-Maximilians-Universit\"at M\"unchen, Geschwister-Scholl-Platz 1, D-80539 M\"unchen, Germany\\
$^{3}$ Technische-Universit\"at M\"unchen, Arcisstr.~21, D-80333 M\"unchen, Germany}

\date{July, 21$^{\text{st}}$ 2016}
\begin{abstract}
Stochastic differential equations describe well many physical, biological and sociological systems, despite the simplification often made in their derivation. Here the usage of simple stochastic differential equations to characterize and classify complex dynamical systems is proposed within a Bayesian framework. To this end, we develop a \textit{dynamic system classifier} (DSC). The DSC first abstracts training data of a system in terms of time dependent coefficients of the descriptive stochastic differential equation. Thereby the DSC identifies unique correlation structures within the training data. For definiteness we restrict the presentation of DSC to oscillation processes with a time dependent frequency $\omega(t)$ and damping factor $\gamma(t)$. Although real systems might be more complex, this simple oscillator captures many characteristic features. The $\omega$ and $\gamma$ timelines represent the abstract system characterization and permit the construction of efficient signal classifiers. Numerical experiments show that such classifiers perform well even in the low signal-to-noise regime. 

\end{abstract}

\maketitle
\section{Introduction}
\label{sec:Introduction}
\subsection{Basic idea}
\label{subsec:basic_idea}
A classification problem for complex dynamic systems might look as follows: For a number of different system classes a few training samples of the evolution of some of the variables for each class are obtained, either by observation or by numerical simulation. Now, different observed systems should be classified with respect to the reference classes. 

In seismology, for example, numerical simulations of
earthquakes are possible, but so time consuming that
for a given observed  seismic event a matching simulation
can rarely be found \cite{1592907, GPR:GPR12161}. This demands for robust methods
to identify characteristic patterns within a limited sample
of simulated waveforms which then can be used to classify
seismic observations with respect to the learned patterns.

In astrophysics, for instance, the field equations of general relativity predict the existence of periodic space time perturbations, so called gravitational waves. These events are generated in violent cosmic events, like the merging of neutron stars and block holes \cite{PhysRevLett.116.061102}, or by various supernova explosions. Unfortunately, it is often computationally not feasible to simulate these processes for all potential initial conditions. A complete catalogue of possible gravitational wave forms is therefore out of reach. Identification and classification of gravitational wave signals need to abstract the few available signal templates in an appropriate fashion, to be able to recognize similar, but not exactly known waveforms in real data. Numerical simulations of supernova explosions exploring parts of the parameter space, showed that one can classify their gravitational wave signatures into three different main types \cite{2008PhRvD..78f4056D}. This should permit to discriminate among these types, if from the few available numerical samples an abstract description can be learned, which can be confronted to real, noisy data. A way to abstract training samples appropriately, permitting one to recognise general characteristics of time lines and features of a system class, is therefore desirable. To this end we propose the usage of simplified stochastic dynamical systems with time dependent coefficients to abstract the time evolution of system classes.  Once the time dependent coefficients are obtained, they serve as a reference for an inexpensive data-system class comparison. We develop both steps, the 'model training', i.e.~the coefficient determination and the 'model selection', i.e.~the classification of an observation into the previous learned system classes. This is done within information field theory (IFT) \cite{2009PhRvD..80j5005E}, a Bayesian inference framework for continuous quantities. 

As periodicity plays a remarkable role in the description of dynamical systems we use the stochastic oscillator equation 
\begin{equation} 
\frac{d^2 x(t)}{dt^2} + \gamma\left(t\right) \frac{dx(t)}{dt} + \omega^2(t) x(t) = \xi(t)\,
\label{eq:oscillator}
\end{equation}
as our baseline model. Here the time dependent frequency $\omega (t)$ and damping factor $\gamma(t)$ characterize each system class and the external force $\xi(t)$ is regarded as a random quantity which generates the variance within each class. $\omega(t)$ and $\gamma(t)$ have to be learned from a training data set of this system class. This system class is then identified by its $\omega(t)$ and $\gamma(t)$, while we assume that the exact realization of $\xi(t)$ is unknown and differs from realization to realization within that class. In essence, $\omega(t)$ and $\gamma(t)$ serve as a fingerprint of the system while $\xi(t)$ plays the role of the random seed generating individual realization. 
The non-stationarity of Eq.~\eqref{eq:oscillator}, induced by the time dependent parameters $\gamma(t)$ and $\omega(t)$ allows us to also model anharmonic system classes which do not show a clear oscillating behavior. 

The adopted stochastic model, here the oscillator in Eq.~\eqref{eq:oscillator}, is intended to be a simplified description of a  much more complex system in reality. We intend to find empirical parameters, here $\omega(t)$ and $\gamma(t)$, which characterize the system class well, without them necessarily being unique, perfectly optimal or even physically interpretable. For this reason, we will be pragmatic when determining the parameters. 
The strategy presented in this work is first to analyze system classes in terms of the class' time-wise varying frequency $\omega(t)$ and damping factor $\gamma(t)$. To do so, we apply a hierarchical Bayesian ansatz within the framework of IFT as this also allows us to simultaneously reconstruct the two point correlation function of $\omega(t)$ and $\gamma(t)$. For each analyzed system class, $\omega(t)$ and $\gamma(t)$ serve as abstract classification signatures to which observational data can be confronted. Consequently, we introduce a Bayesian model comparison approach using these signatures to state the probability that a given observation is explained by one of the learned system classes.

\subsection{Previous work}
\label{subsec:previous_work}

In the last decades, extensive research has been made within the field of estimating parameters of deterministic and stochastic systems, respectively \cite{Mendes01011998, Kushner:1964qv, Jones:1998:EGO:596070.596218, Betro:1991fv, Mockus:1994qf, 23094, Zilinskas:1992qd, Jones:1998cr}. Most attention has been drawn to local and global nonlinear optimization methods. The use of stochastic processes within the global optimization approaches is called \textit{Bayesian global optimization} or the \textit{random function approach}. In a more generic setting general parameter estimation has been performed extensively by maximum likelihood estimations \cite{RSSC:RSSC5148, Baker200550}. In financial mathematics the parameter estimation for stochastic models has been extensively studied in a frequentist maximum likelihood as well as in a Bayesian framework \cite{Johannes:oq}. 

In the past, model selection has primarily been performed by likelihood ratio tests \cite{1101146}. Due to the enormous increase of computational power, Bayesian methods have been coming into use more and more \cite{Girolami20084}. 

The principal component analysis (PCA), emerging from \cite{PCA} has successfully been used to reduce data to be represented by a linear superposition of a few uncorrelated principle components \cite{jolliffe2006principal, bishop06a}.  As a PCA might perform poorly in case the superposition principle is violated, neural networks (NN) for pattern recognition and model selection are used more extensively \cite{1165576}. Despite the success of NN in the domain of pattern recognition \cite{Bishop:1995:NNP:525960} there is still an ongoing discussion how to properly adjust NN to their desired task \cite{4471901, Lawrence:1997nr, JMLR:v15:srivastava14a}. We believe that the dynamic system classifier (DSC) can perform better than a PCA or NN, especially in the field of gravitational wave data analysis. There the quality and number of the training samples is often not sufficient to train a PCA or NN. A PCA based classifier, if trained with sufficiently many templates might be too rigid with respect to systematic errors in the training data set. PCA based classifiers compare linear combinations of rigid templates to the data, while DSC might turn out to be more flexible as it classifies the signal covariance structure. NN might outperform both, PCA and DSC. However, they do not provide easy insight into what features of a signal triggers the classification and typically need bigger training samples. In the following we will show that the DSC can properly distinguish between different system classes using just a few training samples for each class. A comparison of DSC to NN and PCA is left for future works as it will strongly depend on the use case at hand.

\subsection{Structure of this work}
\label{subsec:structure_of_work}
This paper is structured as follows: In Sec.~\ref{sec:IFT_intro} we introduce some basic notation for signal inference in IFT. In Sec.~\ref{sec:inference} we establish the training algorithm to infer the time dependent fields $\omega$ and $\gamma$ from training data. In Sec.~\ref{sec:Model_selection} the inferred fields are used to construct our model selection algorithm. After deriving the whole dynamic system classifier (DSC) algorithm we apply it to a realistic test scenario in Sec.~\ref{sec:algo}. We conclude in Sec.~\ref{sec:conclusion}. 

\section{Inference of fields}
\label{sec:IFT_intro}
\subsection{Basic notation}
First of all, we have to establish some notations and basic assumptions. To do so, we primarily follow the notation used in IFT. 
In this paper we will suppose that we are analyzing a discrete set of data $d= \left( d_1, \dots, d_r\right)^\mathsf{T} \in \mathbb{R}^{\text{r}}$, which may depend on the underlying signal $s: \mathcal{S} \rightarrow \mathbb{R}$. By $\mathcal{S} $ we denote the continuous subspace of a u-dimensional Euclidian space on which the signal is defined. 

From the \textit{principal of minimal updating} \cite{kullback1951} and the \textit{principal of maximum entropy} \cite{PhysRev.106.620} it follows that a Gaussian is the proper probability distribution for a quantity which is only characterized by its first and second momentum, as it often occurs in physical experiments. 
By 
\begin{equation}
\mathscr{G} \left( \phi, \vec{\Phi} \right) = \frac{1}{\vert 2 \pi \vec{\Phi} \vert ^{1/2}} \exp \left( - \frac{1}{2} \phi^{\dagger} \vec{\Phi}^{-1} \phi\right)
\label{eq:Gaussian}
\end{equation} we denote  a multivariate Gaussian probability distribution function of a continuous field $\phi$. $\vert \vec{\Phi}\vert$ denotes the determinant and $\phi^\dagger$ the transposed and complex conjugated $\phi$.  The covariance $\vec{\Phi} = \langle \phi \phi^\dagger \rangle_{\mathscr{G}(\phi, \vec{\Phi})}$ can be regarded as a function of two arguments $\vec{\Phi}(x, y)= \langle \phi(x) \phi(y)^\dagger\rangle$ or as a matrix-like operator $\vec{\Phi}_{xy}= \langle \phi_x\phi_y^\dagger\rangle$, where we introduced the index notations for functions $\psi_x= \psi(x)$. Under the assumptions of a statistical stationary or homogeneous process over a $u$-dimensional space 
\begin{equation}
\vec{\Phi}_{xy}=C(x-y)
\label{eq:homogeneous}
\end{equation}
one can show that the covariance $\vec{\Phi}$ becomes diagonal in Fourier space, 
\begin{equation}
\vec{\Phi}_{kq}= (2 \pi)^{u} \delta(k-q) P_\phi (k)\,.
\label{eq:Power_spec}
\end{equation}
We use $k$ to denote frequencies in the Fourier convention $f_k= \int \, dt\, e^{i kt}f_t$ and $f_t= \int \,\frac{dk}{2\pi}f_k e^{-ikt}$. In Eq.~\eqref{eq:Power_spec} we have also introduced the power spectrum $P(k)$ which is identical to the Fourier transform of $C(x-y)$. 
In order to apply any operator to a field we have to specify the scalar product, which we take as
\begin{equation}
\phi^{\dagger} \psi = \int _{\mathcal{S}} dx\,  \overline{\phi\left(x\right)} \psi\left(x\right) \quad \forall\,  \phi, \psi : \mathcal{S} \rightarrow \mathbb{R}\, .
\label{eq:scalar_product}
\end{equation} 
Stationary Gaussian processes with zero mean are fully determined by their power spectrum. Here we aim at characterizing non-stationary processes, for which Eq.~\eqref{eq:homogeneous} and \eqref{eq:Power_spec} do not hold. We construct a non-stationary process $x(t)$ by Eq.~\eqref{eq:oscillator}, in which a fixed time evolution of $\omega(t)$ and $\gamma(t)$ imprints non-stationary correlation structures onto $x(t)$ for a set of noise $\xi(t)$ realizations. The noise realizations $\xi(t)$ as well as the characteristics of $\omega(t)$ and $\gamma(t)$ of a model class are themselves assumed to be realizations of stationary processes. Nevertheless, $x(t)$ is non-stationary for fixed $\omega(t)$ and $\gamma(t)$. 

\subsection{Signal inference}
\label{subsec:signal_inference}
Informative physical experiments provide data from which an unknown signal of interest can be inferred. Since there might be infinitely many possible signal field configurations leading to the same data set, there is no exact solution to this inference problem. Consequently, we have to use probabilistic data analysis methods to obtain the most plausible signal field including its uncertainty.

Given a set of data $d$, the \textit{posterior} probability distribution $\mathcal{P}(s \vert d)$ describes how probable the signal $s$ is given the observed data $d$. This posterior can be calculated according to Bayes' theorem, 
\begin{equation}
\mathcal{P} ( s\vert d) = \frac{\mathcal{P}( d\vert s) \mathcal{P}\left( s\right)}{\mathcal{P}\left(d\right)}\, ,
\label{eq:Bayes_Theorem}
\end{equation}
which is the quotient of the product of the \textit{likelihood} $\mathcal{P}\left( d\vert s\right)$ and the signal \textit{prior} $\mathcal{P}(s) $ divided by the \textit{evidence} $\mathcal{P}(d)$. The likelihood describes how likely it is to measure the observed data set $d$ given a signal field $s$. It covers all processes that are relevant for the measurement of $d$. The prior characterizes all a-priori knowledge and therefore does not depend on the data itself. 
As we are interested to find the most plausible signal field configuration given the observed $d$, we use the \textit{maximum a posteriori} ansatz (MAP). The outcome of the MAP- Ansatz states the most probable field configuration
\begin{equation}
m=\underset{s}{ \text{argmax}} \{ \mathcal{P}(s \mid d)\}\,.
\end{equation}
In the following sections we will now discuss the likelihood and the prior of the evolution of a stochastic system, which is described by Eq.~\eqref{eq:oscillator}. 

\section{Model Training}
\label{sec:inference}
\subsection{The likelihood of a stochastic differential equation}
\label{subsec:likelihood_SDE}
As outlined, we use an oscillator model with evolving frequency $\omega$ and damping factor $\gamma$ to characterize a class of systems. To ensure strict positive definiteness of the time dependent frequency we parametrize it as $\omega^2(t)= \omega_0^2 e^{\beta\left(t\right)}$ in Eq.~\eqref{eq:oscillator}. Here we choose time units such that $\omega_0=1$. Consequently, the actual fluctuations of the frequency are characterized by $\beta\left(t\right)$. 

\begin{figure}
    \centering
        \begin{tikzpicture}
            [c/.style={circle,minimum size=2em,text centered,thin},
             r/.style={rectangle,minimum size=2em,text centered,thin},
             v/.style={->,shorten >=1pt,>=stealth,thick}, 
             arrow/.style={-latex, shorten >=1ex, shorten <=1ex, bend angle=45}]
            \node(a)at(-1.3,2)[c, draw]{$\beta_t$};
	    \node(b)at(1.2,2)[c, draw]{$\gamma_t$};
	    \node(c)at(0,1.5)[c, draw]{$s$};
	    \node(d)at(0,0)[c, draw]{$\R$};
	    \node(e)at(0,-1.5)[c, draw]{$x_t$};
	    \draw[v](a)--(c);
	    \draw[v](b)--(c);
	    \draw[v](c)--(d);
	    \draw[v](d)--(e);
        \end{tikzpicture}
        \flushleft
        \caption{Hierarchical Bayes model for the key quantities. The two fields $\beta_t$ and $\gamma_t$ form together a signal $s$, defining the response $\R$, with which the system reacts to the driving white noise $\xi$. The application of $\R$ on white Gaussian noise solves Eq.~\eqref{eq:oscillator} and consequently yields to its solution $x=\R\xi$. From sufficient training data vectors $x$ a plausible signal should be inferred in the learning phase.}
        \label{fig:SDE_inference}
\end{figure}
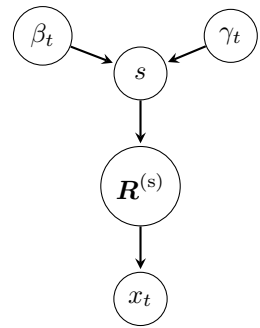

Figure \ref{fig:SDE_inference} shows a Bayesian network to infer the time dependent parameters $\beta_t$ and $\gamma_t$ from the solution $x_t$ of a stochastic differential equation as in Eq.~\eqref{eq:oscillator}. The inferred vector $s=\left(\beta_t, \gamma_t\right)$ will serve as the characteristic signature of a system class in section \ref{sec:Model_selection}. 
To construct this Bayesian inference network we note that the linearity of Eq.~\eqref{eq:oscillator} implies 
\begin{equation}
x = \vec{R} \xi\,, 
\end{equation}
and in scalar product notation implying
\begin{equation}
x_t= \int\, dt'\vec{R}_{tt'} \xi_t'\,, 
\label{eq:x=rxi}
\end{equation}
where we have introduced the response operator $\vec{R}_{tt'}= \vec{R}(t,t')$. The force $\xi_t$ to the oscillator is assumed to be white Gaussian noise, i.e. $\mathcal{P}\left( \xi \vert \vec{\Xi}\right)= \mathscr{G}\left(\xi, \vec{\Xi}\right)$, with a given diagonal and constant covariance $\vec{\Xi}_{tt'} = \delta_{tt'} \tilde{\Xi}$ and the amplitude $\tilde{\Xi}$. Given that the response operator $\vec{R}$ and its inverse $\vec{R}^{-1}$ exist, Eq.~\eqref{eq:x=rxi} can be rewritten as
\begin{equation}
\int\, dt'\, \vec{R}^{-1}_{tt'} x_t= \xi_t\,,
\label{eq:R-1x=xi}
\end{equation}
which allows us to identify the functional form of the response operator $\vec{R}_{tt'}$ by comparing Eq.~\eqref{eq:R-1x=xi} with Eq.~\eqref{eq:oscillator}. Hence $\vec{R}_{tt'}$ is a reformulation of the differentials\footnote{The defining property of a  Dirac delta distribution $\delta(x)$ is
\begin{equation*}
\int_{-\infty}^{\infty} f (  t) \delta ( t-a ) dt \equiv \int_{-\infty}^{\infty} f (t) \delta_{t a} dt= f(a)\,.
\label{dirac_delta}
\end{equation*}
Its derivatives are given by
\begin{equation*}
\int f ( t ) \delta^{\left( n \right) } ( t ) dt \equiv -\int \frac{\partial f(t )}{\partial t} \delta^{( n-1 )} ( t) dt\,.
\label{dirac_delta_derivative}
\end{equation*}} in Eq.~\eqref{eq:oscillator} and is defined as
\begin{equation}
\left(\Rt \right)^{-1} := \delta^{\left(2 \right)}_{tt'}- \gamma(t) \delta^{\left ( 1 \right) }_{tt'} + e^{\beta(t)} \delta_{tt'} \,.
\label{eq:def_response}
\end{equation}
As already used in Eq.~\eqref{eq:def_response} we will from now on refer to the response operator as $\R$ to indicate that it depends on the signal $s$, which is characteristic for a system class. 
With Eq.~\eqref{eq:x=rxi} the likelihood  $\mathcal{P}(x\vert s)$ becomes
\begin{align}
\mathcal{P}(x \vert s)&= \int \mathcal{D}\xi \, \delta(x- \R\xi)\, \mathscr{G}(\xi,\vec{\Xi}) \notag \\
						&= \mathscr{G} \left(x, \R^\dagger \vec{\Xi} \R\right)\,, 
\label{eq:likelihood}
\end{align}
where we have marginalized over all possible realizations of the driving force $\xi_t$. We see that $x_t$ is modeled as a Gaussian random field with a temporarily structured covariance $\vec{X}= \R^\dagger \vec{\Xi} \R$. Thus, $\vec{X}$ is not of the from given by Eq.~\eqref{eq:homogeneous} and \eqref{eq:Power_spec}, respectively and specified a non-stationary process, which is characterized by $s$.

\subsection{Prior assumptions}
\label{subsec:prior}
\begin{figure}[h]
    \centering
        \begin{tikzpicture}
            [c/.style={circle,minimum size=2em,text centered,thin},
             r/.style={rectangle,minimum size=2em,text centered,thin},
             v/.style={->,shorten >=1pt,>=stealth,thick}, 
             arrow/.style={-latex, shorten >=1ex, shorten <=1ex, bend angle=45}]
            \node(a)at(-2,5)[r,text width=3em,draw]{$\alpha_{\beta},\;q_{\beta}$};
            \node(b)at(-.75,5)[r,draw]{$\sigma_{\beta}$};
            \node(c)at(-1.3,3.5)[c,draw]{$\tau_{\beta}$};
            \node(d)at(-1.3,2)[c, draw]{$\beta_t$};
	    \draw[v](a.south)--(c);
	    \draw[v](b.south)--(c);
	    \draw[v](c.south)--(d);	   
        \end{tikzpicture}
        \flushleft
        \caption{The spectral parameters $\alpha_{\beta}$ and $q_{\beta}$ define together with the smoothness enforcing parameter $\sigma_{\beta}$ the prior for the shape of the spectral parameters $\tau_\beta(k)$. Hence the correlation structure of $\beta_t$ is described by $\tau_\beta(k)$.}
        \label{fig:hier_prior}
\end{figure}
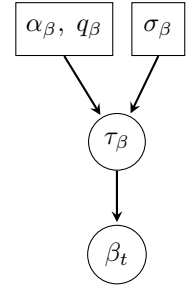

As the likelihood in Eq.~\eqref{eq:likelihood} only describes how $\beta_t$ and $\gamma_t$ are transformed into $x_t$ we need to model our prior knowledge about the time evolution of these parameters. To do so we briefly outline a hierarchical Bayesian prior ansatz, which leads to the so called \textit{critical filter} \cite{2011PhRvD..83j5014E, 2013PhRvE..87c2136O}. Figure \ref{fig:hier_prior} shows this hierarchical parameter model, which will be introduced in the following.  
We assume a priori that $\beta_t$ as well as the $\gamma_t$ parameters obey multivariate Gaussian distributions, 
\begin{align}
\mathcal{P}(\beta_t \vert \vec{\Omega} )&= \mathscr{G}(\beta_t, \vec{\Omega})\, , \\
\mathcal{P}(\gamma_t \vert \vec{\Gamma} )&= \mathscr{G}(\gamma_t, \vec{\Gamma})\,.
\label{eq:Field_Gaussian} 
\end{align}
The covariances $\vec{\Omega}$ and $\vec{\Gamma}$ describe the strength of the temporal correlations of $\beta$ and $\gamma$, respectively and thus the smoothness of their fluctuations. A convenient parametrization of the covariances can be found, if our knowledge of the parameters does a priori not single out any instance; i.e. correlations only depend on time intervals. This is equivalent to assume $\beta_t$ and $\gamma_t$ to be statistically stationary. Under this assumption, $\vec{\Gamma}$ and $\vec{\Omega}$ are diagonal in the Fourier space representation, see Eq.~\eqref{eq:Power_spec}. 

To keep notations simple, we will only show the calculations for $\vec{\Omega}$ in full detail since the ones for $\vec{\Gamma}$ can be performed analogously. Later we will see that the priors for $\beta_t$ and $\gamma_t$ only differ by hyperprior-parameters. 

Under the assumption of statistical stationarity, we choose the following ansatz for the covariance: 
\begin{equation}
\vec{\Omega}= \sum_k e^{\tau_\beta(k)} \vec{\Omega}_k\,
\label{eq:basis_prior}
\end{equation}
Here $\tau(k)$ are spectral parameters, determining the power spectrum $P_\beta (k)$ and $\vec{\Omega}_k$ are projection operators onto spectral bands, with approximately identical power spectrum values. By $\vec{\Omega}_k^{-1}$ we refer to the pseudo inverse of the projection operators.
 Since the covariances $\vec{\Gamma}$ and $\vec{\Omega}$ are unknown, one has to introduce another prior for them, i.e. a prior for properly describing the spectral parameters $\tau (k)$, for each $\beta$ and $\gamma$. As the power spectrum's shape might span over several orders of magnitude, this requires a logarithmically uniform prior for each element of the power spectrum and a uniform prior $\mathcal{P}_{\text{un}}$  for each spectral parameter $\tau_k$, respectively. In accordance with \cite{2011PhRvD..83j5014E, 2010PhRvE..82e1112E} we therefore initially assume inverse-Gamma distributions for the individual elements
\begin{equation}
\mathcal{P}_{\text{un}}\left( e^{\tau_k} \vert \alpha_k,\,  q_k\right) = \frac{q_k^{\alpha_k-1}}{\Gamma\left(\alpha_k-1\right)}e^{-\left(\alpha_k \tau_k + q_k e^{-\tau_k}\right)}
\end{equation}
and hence
 \begin{align}
 \mathcal{P}_{\text{un}}\left( \tau_k \vert \alpha_k,\,  q_k\right) &=  \frac{q_k^{\alpha_k-1}}{\Gamma\left(\alpha_k-1\right)}\notag \\
 &\quad \times e^{-\left(\alpha_k \tau_k + q_k e^{-\tau_k}\right)}
  \left \vert \frac{de^{\tau_k}}{d \tau_k}  \right \vert 
 \label{eq:inverse_gamma}
 \end{align}
 where $\alpha_k$ and $q_k$ denote shape and scale parameters for the spectral hyperpriors, and $\Gamma$ the Gamma function. If $\alpha_k \rightarrow 1$ and $q_k \rightarrow 0 \, \forall k$, the inverse Gamma distribution becomes asymptotically flat on a logarithmic scale. In practice, $q_k \, \ge 0$ provide lower limits for the power spectra, which lead to a more stable inference algorithm.  
Note that the variations of $\alpha_k$ and $q_k$ with $k$ can be used to model specific spectral prior knowledge. However, in the absence of such knowledge, these will get the same values $\alpha_k= \alpha$, $q_k= q=\text{const.}$. 

Until now we have only addressed each individual element of the power spectrum separately, but empirically we know that many power spectra do not exhibit strong fluctuations on neighboring Fourier scales. It is therefore natural for the spectrum to request some sort of smoothness. To enforce this behavior, we further incorporate a spectral smoothness prior $\mathcal{P}_{\text{sm}}$ \cite{2011PhRvD..83j5014E, 2013PhRvE..87c2136O}. This spectral smoothness prior is based on the second logarithmic derivative of the spectral parameters $\tau$. Up to a normalization constant $\mathcal{P}_{\text{sm}}$ can be written as 
 \begin{equation}
 \mathcal{P}_{\text{sm}}\left(\tau \vert \sigma \right) \propto \exp \left(-\frac{1}{2} \tau^{\dagger} \vec{T} \tau\right) 
 \label{eq:Prior_smoothness}
 \end{equation}
 with 
 \begin{equation}
\tau^{\dagger} \vec{T} \tau = \int d \left( \log k \right) \frac{1}{\sigma^2}\left( \frac{ \partial^{2} \tau_k}{\partial \left(\log k\right)^{2}} \right)^{2}\,,
\label{eq:operator_smoothness}
\end{equation}
which is punishing deviations from any power-law behavior of the power spectrum. The strength of the punishment is encoded by $\sigma$.  In total, the resulting prior for the spectral parameters $\tau$ is given by the product of the priors discussed above
\begin{equation}
\mathcal{P}\left(\tau \vert \alpha,\, q, \sigma \right) \propto \mathcal{P}_{\text{sm}}\left( \tau \vert \sigma\right)\,  \prod_k \mathcal{P}_{\text{un}}\left(\tau_k \vert \alpha_k,\, q_k  \right) \,,
\label{eq:criticial_filter_prior}
\end{equation}
with its three given quantities  $\alpha_k, q_k$ and $\sigma$.

With this hierarchical Bayesian model we are able to state the posterior:
\begin{align}
\mathcal{P} (s,  \tau_\beta, \tau_\gamma &\vert x_t, \alpha_{\omega, \gamma}, q_{\omega, \gamma}, \sigma_{\omega, \gamma}) = \notag \\
 & \frac{ \mathcal{P}( x_t \vert s)}{\mathcal{P}( x_t)}  \prod_{i \in \{\beta, \gamma \} }  \mathcal{P}( i \vert \tau_{i}) \mathcal{P}( \tau_{i} \vert \alpha_{i},\, q_{i},\, \sigma_{i})
\label{eq:full_posterior}
\end{align}

\subsection{Parameter reconstruction}
Now we are seeking for the most probable parameter configurations of $\beta_t$ and $\gamma_t$ given the training data $x_t$. Due to the complexity of the posterior, given by Eq.~\eqref{eq:full_posterior}, it is virtually impossible to solve this problem analytically. Numerical searches through the parameter space using Markov Chain Monte Carlo methods \cite{Metropolis:1949:MCM, metropolis53} are possible, as similar problems have been solved by advanced sampling techniques \cite{2004PhRvD..70h3511W, 2010MNRAS.407...29J, 2010MNRAS.406...60J, 2013ApJ...779...15J}. Given that the employed stochastic partial differential equation is just an empirical guess we pragmatically use here the numerically and computationally cheaper MAP-ansatz. It is not necessary to find the best parameters, a good classification will usually suffice.
Rather than maximizing the posterior it is thereby convenient to define the negative logarithm of the posterior $\mathcal{P}\left(s, \tau_\beta, \tau_\gamma\vert x\right)$ as the \text{information Hamiltonian}  
 \begin{align}
 \begin{split}
&\mathscr{H}\left( s, \tau_\beta, \tau_\gamma\vert x_t \right) = -\log \mathcal{P}\left( s, \tau_\beta, \tau_\gamma\vert x_t \right) \\  
&  = \log\left( \det\left[\R\right]\right)\\
  &\quad + \frac{1}{2} x_t^{\dagger} \left(\R^{-1}\right)^{\dagger} \Xi^{-1} \R^{-1} x_t \\
  &\quad + \frac{1}{2} \log \left(\det \left [\vec{\Omega}\right]\right) +\frac{1}{2} \beta_t^{\dagger} \vec{\Omega}^{-1} \beta_t\\
  &\quad + \frac{1}{2} \log\left( \det \left[ \vec{\Gamma}\right]\right)+\frac{1}{2} \gamma_t^{\dagger} \vec{\Gamma}^{-1} \gamma_t \\
 &\quad+ \sum_{i \in \{\beta_t,\, \gamma_t\}} \left( \alpha_i-1\right)^{\dagger}\tau + q_i^{\dagger} e^{-\tau
 _i} + \frac{1}{2} \tau_i^{\dagger} \vec{T} \tau_i\, , \\
 &\quad +H_0\, ,
 \end{split}
 \label{eq:Hamiltonian}
\end{align}
where we have absorbed all terms constant in $\beta_t$, $\gamma_t$, $\tau_\beta$, and $\tau_\gamma$ into $H_0$. 

By this reformulation the MAP solution is now seeking for the minimum of Eq.~\eqref{eq:Hamiltonian}. This minimum may be found by taking the derivative of the information Hamiltonian with respect to $\beta_t$, $\gamma_t$, $\tau_\gamma$, and $\tau_\beta$, respectively and equating them with zero. This yields four implicit equations. The minimum we are seeking for may be found by an iterative downhill algorithm such as the steepest descent. 
To better understand the MAP solution we focus on the resulting filtering formulas of this ansatz. The ones for the frequency and damping factor evolution read 
\begin{align}
\begin{split}
\frac{\partial \mathscr{H}}{\partial \beta_t} \Bigg\vert_{\beta=\beta_{\mathrm{min}}} &= \tr \left [ \R^{-1} \frac{\partial \R}{\partial \beta_t} \right ]\\
	& \quad + \frac{1}{2} x_t^{\dagger} \left( \frac{\partial \R^{-1}}{\partial \beta_t}\right)^{\dagger} \vec{\Xi}^{-1} \R^{-1} x_t \\
	& \quad +  \frac{1}{2} x^{\dagger} \left(\R^{-1}\right)^{\dagger} \vec{\Xi}^{-1} \frac{\partial\R^{-1}}{\partial \beta_t}x_t \\
	& \quad + \vec{\Omega}^{-1} \beta_t \\
	& \overset{!}{=}0 
\end{split} 
\label{eq:Hamiltonian_D_w}
\end{align}
and
\begin{align}
\begin{split}
\frac{\partial \mathscr{H}}{\partial \gamma_t} \Bigg\vert_{\gamma_t=\gamma_{\mathrm{min}}} &= \tr \left [ \R^{-1} \frac{\partial \R}{\partial \gamma_t} \right ]\\
	& \quad + \frac{1}{2} x_t^{\dagger} \left( \frac{\partial \R^{-1}}{\partial \gamma_t}\right)^{\dagger} \vec{\Xi}^{-1} \R^{-1} x_t \\
	& \quad +  \frac{1}{2} x^{\dagger} \left(\R^{-1}\right)^{\dagger} \vec{\Xi}^{-1} \frac{\partial\R^{-1}}{\partial \gamma_t} x_t \\
	& \quad + \vec{\Gamma}^{-1} \gamma_t \\
	& \overset{!}{=}0\,.
\end{split}
\label{eq:Hamiltonian_D_y}
\end{align}
While the filter formula for the power spectra of $\beta_t$ is 
\begin{align}
\begin{split}
\frac{\partial \mathscr{H}}{\partial \tau_{\beta}} \overset{!}{=}0 \rightarrow e^{\tau_\beta}&= \frac{q_{\beta} + \frac{1}{2} \left( \tr \left [ \beta_t \beta_t^{\dagger} \vec{\Omega}_k^{-1}\right] \right)_k}{\left( \alpha_{\beta} -1 \right) + \frac{1}{2} \left( \tr \left [ \vec{\Omega}_k \vec{\Omega}_k^{-1}\right]\right)_k + \vec{T} \tau_{\beta}} 
\end{split} 
\label{eq:P_D_w}
\end{align}
and the one for $\gamma_t$ is
\begin{align}
\begin{split}
\frac{\partial \mathscr{H}}{\partial \tau_{\gamma}}  \overset{!}{=}0 \rightarrow e^{\tau_\gamma}&= \frac{q_{\gamma} + \frac{1}{2} \left( \tr \left [ \gamma_t \gamma_t^{\dagger} \vec{\Gamma}_k^{-1}\right] \right)_k}{\left( \alpha_{\gamma} -1 \right) + \frac{1}{2} \left( \tr \left [ \vec{\Gamma}_k \vec{\Gamma}_k^{-1}\right]\right)_k + \vec{T} \tau_{\gamma}}  \,.
\end{split} 
\label{eq:P_D_y}
\end{align}
The filtering formulas in Eq.~\eqref{eq:P_D_w} and \eqref{eq:P_D_y} have previously been derived Refs.~\cite{2014AIPC.1636...49E, 2013PhRvE..87c2136O}. Due to the construction of the hierarchical Bayesian parameter model the covariance structures of $\beta_t$ and $\gamma_t$ get also inferred from $x_t$. The spectral shapes of $\vec{\Omega}$ and $\vec{\Gamma}$ are only constrained by Eq.~\eqref{eq:criticial_filter_prior}. 

In total, the model learning phase leads to a restriction of possible signals to the set 
\begin{equation}
\mathscr{S}= \{s_1, s_2, \dots s_n\}\, , 
\end{equation}
which is finite in case of a finite number of considered system classes. 

\section{Model selection}
\label{sec:Model_selection}
So far we only faced the inverse problem to reconstruct time dependent parameters, such as the frequency $\beta_t$ and the damping factor $\gamma_t$ including their their power spectra from an oscillator driven by a stochastic force. From now on we will extend our model to a measurement scenario, involving a measurement response $\vec{R}_{\text{obs}}$ and additive Gaussian measurement noise $n \hookleftarrow \mathscr{G} \left( n, \vec{N} \right)$, with related covariance $\vec{N}$. Consequently the data model is now given by 
\begin{equation}
\vec{d} = \vec{R}_{\text{obs}} x + n = \vec{R}_{\text{obs}} \R \xi+ n\,.
\label{eq:realistic_data}
\end{equation}
where $\R$, including $s=(\beta_t, \gamma_t)$, serves as an abstract operator to classify, identify and distinguish between different physical systems $s_1, s_2, \dots$. Hence each $s$ acquires its $\R$ from training data $x(t)$ of its specific physical system $s_i$, according to the previously described algorithm. To state the probability of a model $s_i$ in the set of possible signals $\mathscr{S}$ given the observed data we again use Bayes' theorem
\begin{equation}
\mathcal{P}\left(s_i\vert \mathscr{S}, \vec{d}\right)= \frac{\mathcal{P}\left( \vec{d}\vert \mathscr{S}, s_i\right) \mathcal{P}\left(s_i\vert \mathscr{S}\right)}{\mathcal{P}\left(\vec{d}\right)}\, .
\label{eq:model_decision_posterior}
\end{equation}
The involved likelihood turns out to be
\begin{align}
\mathcal{P}( d \vert s_i) &=   \int \mathcal{D}x\, \mathcal{P} ( d \vert x) \mathcal{P} (x\vert s_i) \notag \\
							&=\int \mathcal{D}x\, \, \mathscr{G} ( d-\vec{R}_{\text{obs}} x,N) \notag \\ 
							& \qquad \times \mathscr{G} (x, \R^\dagger \Xi \R) \notag\\
							&\propto \frac{1} {\sqrt{\vert D \vert }} \exp \left( \frac{1}{2}j^{\dagger} D j \right )  			
\label{eq:likelihood_model_selection}
\end{align} 							
with 
\begin{equation}
j= \R^{\dagger} \vec{R}_{\text{obs}}^{\dagger}N^{-1} d 
\end{equation}
and 
\begin{equation}
D^{-1}=  \R^{\dagger}\vec{R}_{\text{obs}}^{\dagger}N^{-1}\vec{R}_{\text{obs}}\R + \vec{\Xi}^{-1}.
\end{equation}
With this equation one is able to calculate the model posterior, Eq.~\eqref{eq:model_decision_posterior}, and to state the most propable model $s_i$. 

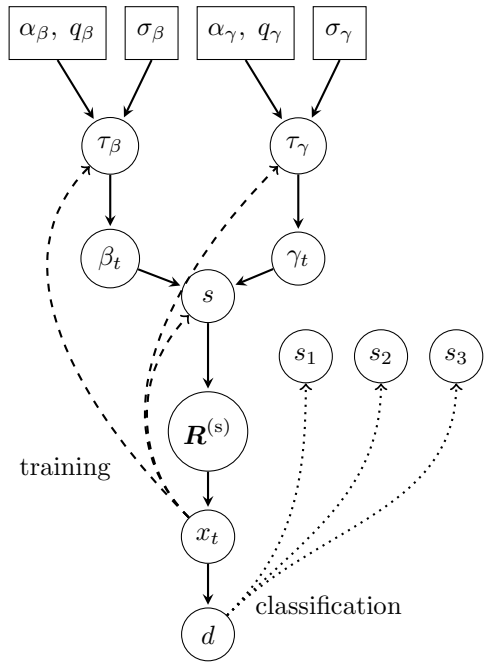
\begin{figure}
    \centering
        \begin{tikzpicture}
            [c/.style={circle,minimum size=2em,text centered,thin},
             r/.style={rectangle,minimum size=2em,text centered,thin},
             v/.style={->,shorten >=1pt,>=stealth,thick}, 
             arrow/.style={-latex, shorten >=1ex, shorten <=1ex, bend angle=45}]
            \node(a)at(-2,5)[r,text width=3em,draw]{$\alpha_{\beta},\;q_{\beta}$};
            \node(b)at(-.75,5)[r,draw]{$\sigma_{\beta}$};
            \node(c)at(-1.3,3.5)[c,draw]{$\tau_{\beta}$};
            \node(g)at(-1.3,2)[c, draw]{$\beta_t$};
	    \node(d)at(.5,5)[r,text width=3em,draw]{$\alpha_{\gamma},\;q_{\gamma}$};
	    \node(e)at(1.75,5)[r,draw]{$\sigma_{\gamma}$};
	    \node(f)at(1.2,3.5)[c,draw]{$\tau_{\gamma}$};
	    \node(h)at(1.2,2)[c, draw]{$\gamma_t$};
	    \node(ha)at(0,1.5)[c, draw]{$s$};
	    \node(h1)at(1.3,.7)[c,draw]{$s_1$};
	    \node(h2)at(2.3,.7)[c,draw]{$s_2$};
	    \node(h3)at(3.3,.7)[c,draw]{$s_3$};
	    \node(i)at(0,-.3)[c, draw]{$\R$};
	    \node(j)at(0,-1.7)[c, draw]{$x_t$};
	    \node(k)at(0,-3)[c, draw]{$ d$};
	    \node(tt)at(-1.9, -.8){training};
	    \node(cc)at(1.6, -2.6){classification};
	    \draw[v](a.south)--(c);
	    \draw[v](b.south)--(c);
	    \draw[v](d.south)--(f);
	    \draw[v](e.south)--(f);
	    \draw[v](c)--(g);
	    \draw[v](f)--(h);
	    \draw[v](g)--(ha);
	    \draw[v](h)--(ha);
	    \draw[v](i)--(j);
	    \draw[v](ha)--(i);
	    \draw[v](j)--(k);
	    \draw[->, dashed, thick](j) to [out=135, in=225](ha);
	    \draw[->, dashed, thick](j) to [out=135, in=225](f);
	    \draw[->, dashed, thick](j) to [out=135, in=225](c);
	    \draw[->, dotted, thick, label=$?$](k) to [out=45, in=270](h1);
	    \draw[->, dotted, thick](k) to [out=45, in=270](h2);
	    \draw[->, dotted, thick](k) to [out=45, in=270](h3);	   
        \end{tikzpicture}
        \flushleft
        \caption{Complete Bayesian network and algorithm. The solid arrows show graphical steps by the probabilistic dependencies. The top four boxes indicate the hyper parameters $\alpha_{\beta/ \gamma}$, $q_{\beta/ \gamma}$, and $\sigma_{\beta/ \omega}$. Below them follow the spectral parameters $\tau_{\beta/ \gamma}$ and the tuple of the frequency $\beta_t$ and the damping factor $\gamma_t$, which form together the signal $s$.  Each system class $i$ is thereby characterized by its $s_i$. Beneath that comes the response operator $\R$, the training data $x_t$, and finally the observed data, according to Eq.~\ref{eq:realistic_data}. The dashed lines on the left hand side of the figure display the workflow of the algorithm to learn the time evolution of $\beta_t$ and $\gamma_t$, including their power spectra from the training data set. On the right hand side the dotted lines display the workflow of the algorithm to state which of the previously learned training data set explains the observed data $d$ best.}
        \label{fig:graphical_model}
\end{figure}

Figure \ref{fig:graphical_model} shows an overview of the suggested hierarchical Bayesian decision algorithm. Given the hyper parameters, the algorithm first learns the frequency $\beta_t$ and damping factor $\gamma_t$ evolution from each training data set. The logarithmic power spectrum $\tau$ for $\beta_t$ as well as for $\gamma_t$ can be regarded as a set of nuisance parameters that get reconstructed from the data to properly infer the parameters of interest $\beta_t$ and $\gamma_t$.  After $s_i=(\beta_t, \gamma_t)$ was  learned for a class $i$, it serves as an abstract characteristic for this model. With the knowledge of $s_i$ the algorithm is able to state how probable the previously learned model $i$ would have caused the observed data $d$. This then serves as a proxy probability for the system classification. I

\section{Dynamic system classifier algorithm}
\label{sec:algo}
Inferring time dependent fields, such as a time-wise varying frequency $\beta_t$ and damping factor $\gamma_t$, from a stochastically driven oscillator is a non-trivial task. The reliability of the dynamic system classifier (DSC) algorithm strongly depends on the successful and proper reconstruction of $\beta_t$ and $\gamma_t$ as they serve as classifiers. 
To show the principal capabilities of our suggested algorithm we will first discuss the algorithm to infer time-dependent parameters of a dynamic system driven by a stochastic force as described in Sec.~\ref{sec:inference}. Subsequently, we use the inferred parameters to test the performance of the model selection algorithm as described in Sec.~\ref{sec:Model_selection}. 
\subsection{Model learning algorithm}
\subsubsection{Numerical application}
The information Hamiltonian, Eq.~\eqref{eq:Hamiltonian}, is a scalar quantity defined over the configuration space of possible model parameter evolutions. In addition to the parameters $\beta_t$ and $\gamma_t$, the two spectral parameters $\tau_\beta$ and $\tau_\gamma$ also need to get inferred from a single system trajectory $x(t)$.  

Hence, the algorithm faces an underdetermined inverse problem, which is also reflected in the possibility of local-minima of the non-convex Hamiltonian. Ultimately the complexity of this inverse problem goes back to the generally highly non-linear entanglement between the two parameters $\beta_t$ and $\gamma_t$.  To overcome this problem we strongly advise to analyze as many realizations ($x= x_1, x_2, x_3, \dots x_l$) of the same system as possible. In section \ref{subsec:Numerical_application} we will discuss in more detail how many data realizations are necessary for an appropriate reconstruction of the parameters. 
The training part of the DSC-algorithm is based on an iterative optimization scheme, where certain parts of the problem get alternatively optimized instead of the whole problem simultaneously. To some degree the optimization results are sensitive to the starting values due to the non-convexity of the considered Hamiltonian. However, remaining degeneracies between $\beta_t$ and $\gamma_t$ after exploiting sufficient large training data sets are irrelevant, as these do not strongly discriminate between the members of the training set of a given system class. \\ Based on our experience with variations of the DSC-algorithm we propose the following scheme:

\begin{enumerate}
\item Initialize the algorithm with naive values, such as $\beta_t = \gamma_t =0$ and $\tau_k= \text{const.} \, \forall k$.  
\item Infer $\beta_t$ and $\gamma_t$ via an iterative downhill algorithm, such as steepest descent using the information Hamiltonian Eq.~\eqref{eq:Hamiltonian}, as well as its gradient Eq.~\eqref{eq:Hamiltonian_D_w}  and Eq.~\eqref{eq:Hamiltonian_D_y}. A more sophisticated minimization scheme, such as non-linear conjugate gradient, is conceivable to speed up the algorithm but it would require the full Hessian of Eq.~\eqref{eq:Hamiltonian}.  Multiple test runs have shown that it is  sufficient to evaluate a simplified Hamiltonian
\begin{align}
\begin{split}
&\mathscr{H}\left(s \vert x_1, x_2, \dots x_l \right) \propto \notag \\
&  \frac{1}{2} \sum_{i=1 , \dots }\,  x_i^{\dagger} \left(\left[\R\right]^{-1}\right)^{\dagger} \Xi^{-1} \left(\R\right)^{-1} x_i \notag \\
 &\quad + \frac{1}{2} \log \left(\det \left [\vec{\Omega}\right]\right) +\frac{1}{2} \beta_t^{\dagger} \vec{\Omega}^{-1} \beta_t \notag \\
 &\quad + \frac{1}{2} \log\left( \det \left[ \vec{\Gamma}\right]\right)+\frac{1}{2} \gamma_t^{\dagger} \vec{\Gamma}^{-1} \gamma_t \,,
\end{split}
\end{align}
and its corresponding gradient. The simplified Hamiltonian neglects in particular $\log\left( \det\left[\R\right]\right)$ as it appears in Eq.~\eqref{eq:Hamiltonian}. In contrast to the diagonal covariance matrixes $\vec{\Omega}$ and $\vec{\Gamma}$ the evaluation of the determinant of $\R$ is computationally time consuming due to its complex structure. Numerical experiments with and without $\det \R$ did not show a significant importance of this term. 
\item Use Eq.~\eqref{eq:P_D_w} and Eq.~\eqref{eq:P_D_y}, respectively, to update the priors $\vec{\Omega}$ and $\vec{\Gamma}$. 
\item Repeat step 2 and 3 until convergence. This iterative scheme will take a few cycles until the algorithm has reached its desired convergence level. 
\end{enumerate} 

The spaces of possible parameter configurations of $\beta_t$ and $\gamma_t$ are huge. Consequently, it seems impossible to judge whether the algorithm has converged into the desired global minimum or some local minimum. It might also happen that the reconstructed fields display features which are originally caused by the exciting force $\xi$ and not by the frequency and damping factor itself. These problems can be reduced by the above demonstrated joint analysis of multiple data realizations, as we  see in the following, where we discuss the numerical tests of the optimization scheme.

\subsubsection{Numerical tests}
\label{subsec:Numerical_application}
\begin{figure}
	\begin{center}
	\includegraphics[width=.5 \textwidth]{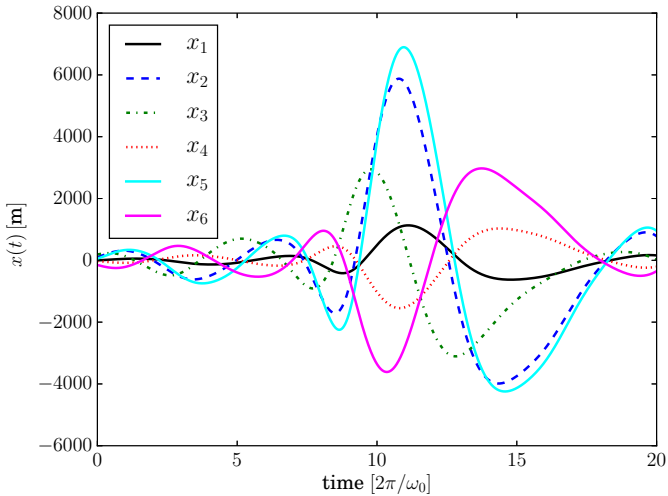}
	\end{center}
	\caption{Six wave realizations according to Eq.~\eqref{eq:x=rxi} for the same model class, i.e. all $x_t$ share the same $\gamma_t$ and $\beta_t$. The amplifying force $\xi$ was drawn from $P\left( \xi \vert \vec{\Xi}\right)= \mathscr{G}\left(\xi, \vec{\Xi}\right)$, with a given constant covariance $\tilde{\Xi} = 5$. }
	\label{fig:data_realizations} 
\end{figure}

To test the performance of the DSC algorithm we applied it to simulated but realistic training data sets (see Fig.~\ref{fig:data_realizations}). The initial conditions, $x(0)$ and $dx(0) /dt$ were set to random values, drawn from a zero centered Gaussian distribution with variance one. This data might represent physical systems whose frequency and damping factor are changing over time. For example, in astrophysics one could expect such a behavior from a gravitational wave burst caused by a supernova \cite{2008PhRvD..78f4056D, 1982A&amp;A...114...53M}. 

In the following tests we used a regular grid with $10^4$ pixels and the signal inference library $\textsc{Nifty}$ \cite{2013A&amp;A...554A..26S} to implement the algorithm. Fig. \ref{fig:data_realizations} shows six realizations of the same simulated system class. This means that the wave realizations, calculated according to Eq.~\eqref{eq:x=rxi} with $\tilde{\Xi}=5$, share the same $\beta_t$ and $\gamma_t$. Note that the waves displayed in Fig.~\ref{fig:data_realizations} are not just rescaled versions of the same wave template. 
For the described MAP reconstruction we used $\alpha=1$, $q= 10^{-30}$, and $\sigma=2$ for both $\gamma_t$ and $\beta_t$.
\begin{figure}
\centering
	\begin{overpic}[width=.48\textwidth]{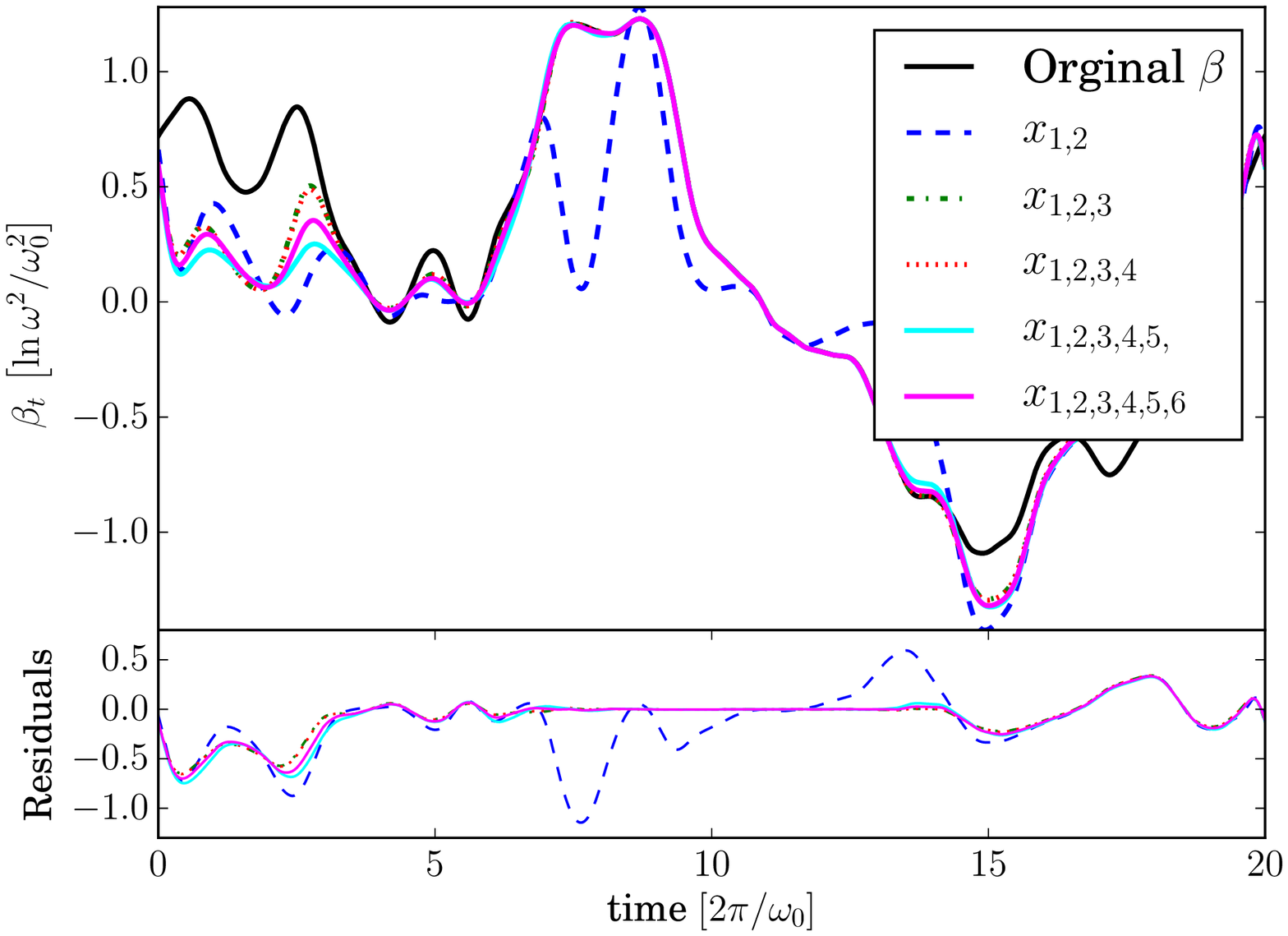} \put(13.5,76.5){(a)} \end{overpic} \\ 
	\vspace{.5cm}
	\begin{overpic}[width=.48\textwidth]{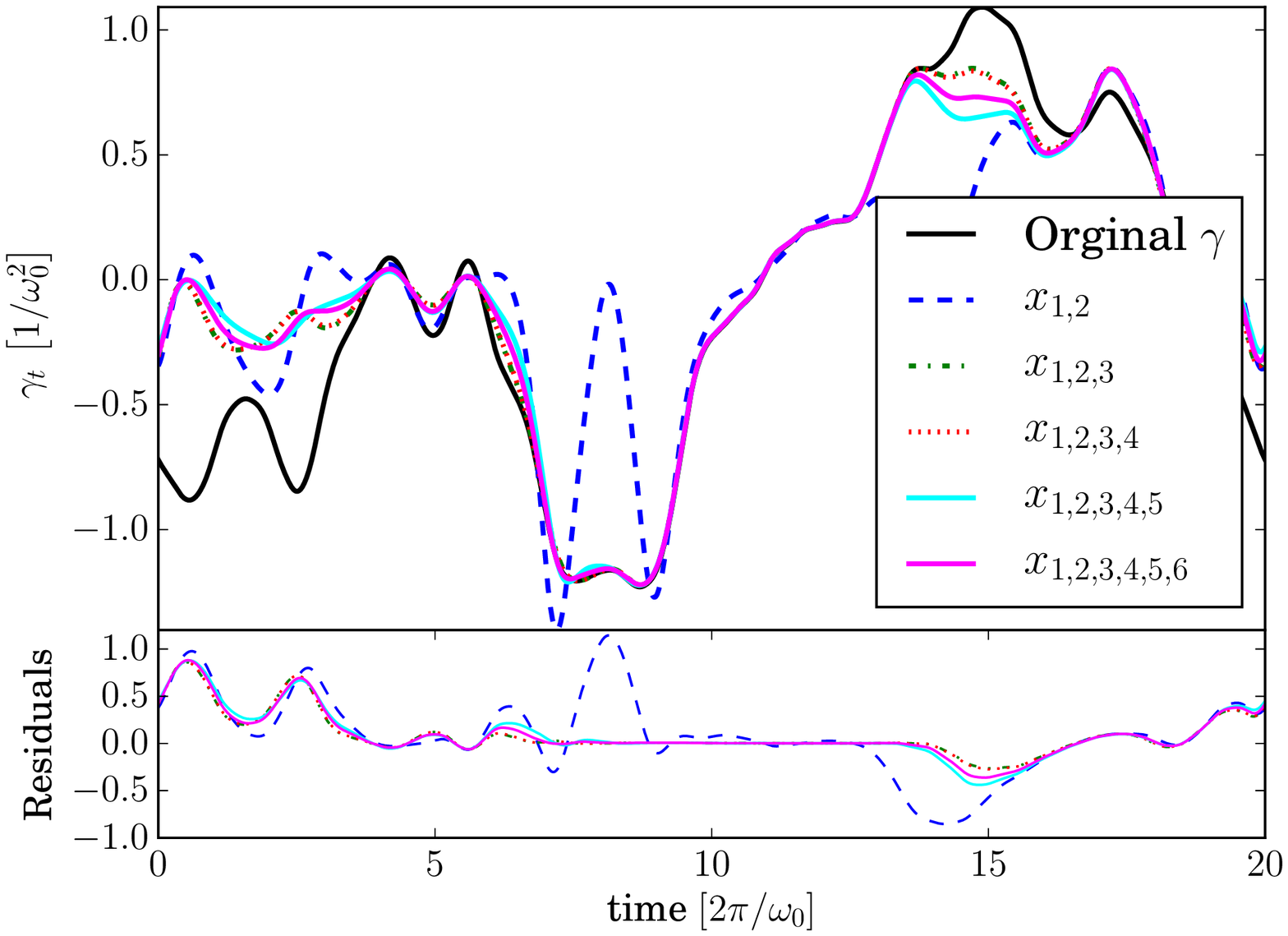} \put(13.5,76.5){(b)} \end{overpic} \\
	\flushleft
    \caption{Panel (a) shows the original $\beta_t$ as well as its reconstruction for different numbers of data realizations $x_t$ ranging from two $(x_{1,2})$ up to six $(x_{1,2,3,4,5,6})$. The residuals between the original $\beta_t$ and their reconstructions are also shown. Panel (b) shows the same for $\gamma_t$ and its reconstructions. One needs at least three $x_t$ to get a proper reconstruction of the fields. Otherwise, the reconstruction shows too many features imprinted by the driving white noise.}
\label{fig:w_y_reconstruction}
\end{figure} 
In Fig. \ref{fig:w_y_reconstruction} the inferred parameters, $\beta_\text{rec}$ and $\gamma_\text{rec}$ including their residuals, as well as the original parameter evolution, $\beta_t$ and $\gamma_t$, are shown. Due to the ahead mentioned degeneracy of the inverse problem we expected that a few wave realizations are needed in order to get reliable reconstructions of the parameters. This is clearly visible in the outcome shown in Fig.~\ref{fig:w_y_reconstruction}, as the residuals between original and reconstructed parameters are largest if one only uses two timelines. Consequently, one needs multiple wave realizations from a system class to get a proper reconstruction and to prevent overfitting of the classe's parameters $\beta_t$ and $\gamma_t$. However, these two parameters are not intended to describe the exact evolution of the system. They only serve as a abstract and descriptive fingerprint of a system class, which is constructed on the basis of Eq.~\eqref{eq:oscillator}.
In case the wave realizations $x(t)$ of a system class  do not provide sufficient information, i.e. have only very small amplitudes, the inference problem becomes more and more degenerated. In this case the reconstructed $\beta_{\text{rec}}$ and $\gamma_{\text{rec}}$ do not reproduce precisely the original ones used in our test to generate the simulated waves. Nonetheless this degeneracy does not destroy the performance of the DSC-algorithm because what counts for the model decision algorithm is the ability of the two parameters to represent the covariance structure for a model class and not whether the parameters are as those generating the timelines of the model classes. 
\begin{figure}
	\includegraphics[width=.48\textwidth]{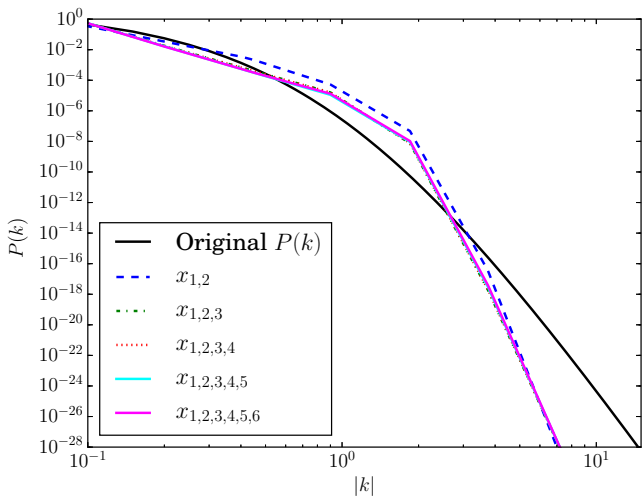}
	\caption{The original power spectrum of $\beta_t$ as well as its reconstruction for
	different numbers of data realizations $x_t$ used in its reconstruction.}
	\label{fig:Prior_rec}
\end{figure}
In Figure \ref{fig:Prior_rec} the inferred power spectrum of $\beta_t$ is shown together with the original one. The power spectrum at small $\vert k\vert$, which corresponds to large-scale correlations, is well reconstructed. In contrast, the power spectrum at large $\vert k \vert$ is underestimated, which may have various reasons. Small short term variations in $\gamma_t$ are nearly indistinguishable from random noise variations in $\xi$. To better reconstruct the small-scale correlation function many more realizations of the system class would be needed. 

In summary we conclude that the presented algorithm can reproduce time dependent parameters from a stochastic differential equation. In the next section we will use the inferred parameters, $\beta_\text{rec}$ and $\gamma_\text{rec}$  to discriminate between different models $s_i$.

\subsection{Performance of model selection algorithm}
\begin{figure}
\centering
	\begin{overpic}[width=.48\textwidth]{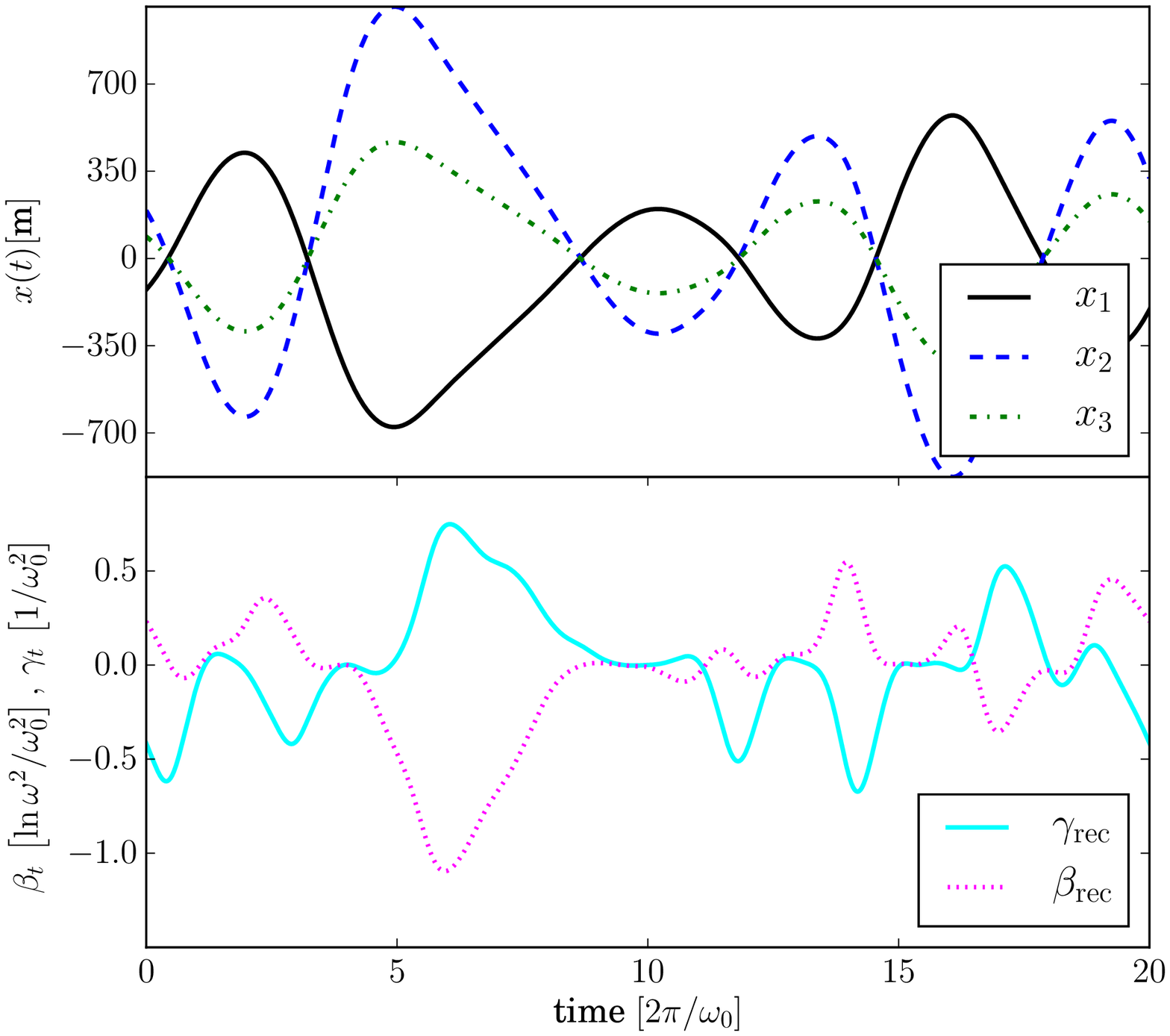} \put(13.5,91.5){(a)} \end{overpic} \\ 
	\vspace{.5cm}
	\begin{overpic}[width=.48\textwidth]{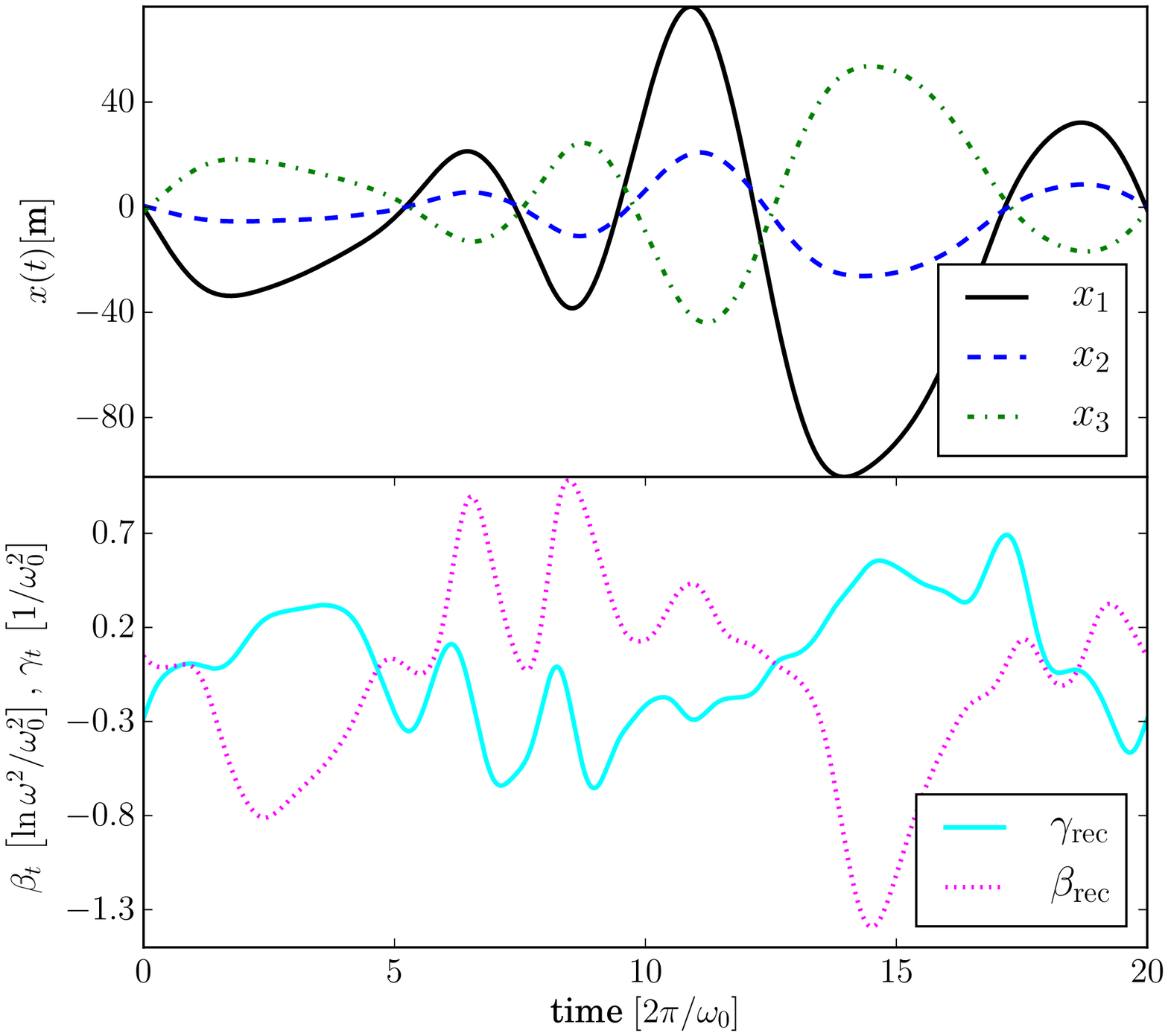} \put(13.5,91.5){(b)} \end{overpic} \\
	\flushleft 
	\caption{In panel (a) the training data set of $s_2$ with its reconstructed parameters, $\omega_{\mathrm{rec}}$ and $\gamma_{\mathrm{rec}}$ is shown. Panel (b) shows the same for $s_3$. }
\label{fig:data}
\end{figure}

To show that $\beta_t$ and $\gamma_t$ can indeed serve as an abstract fingerprint of classes we trained our algorithm with three different training sets, generated from three different models, specified by the parameters $s_1, s_2$, and $s_3$. Each of them had different $\beta_t$ and $\gamma_t$, and for each model we used three wave realizations using the $s_i$ to train the classificator. By $s_1$ we refer to the training data set in Fig.~\ref{fig:data_realizations}, by $s_2$ and $s_3$ we refer to the training data sets shown in Fig.~\ref{fig:data}.  
It is a trivial task to distinguish between the models $s_{1,2,3}$, if the means of $\beta_t$ and $\gamma_t$ differ by several orders of magnitude. To avoid such trivial situations, all $\beta_t$'s and $\gamma_t$'s were drawn from the same power spectrum
\begin{equation}
P\left(k\right) = \frac{42}{\left(1+\vert k\vert \right)^{12}}\,.
\end{equation}

In a second step we generated one noisy data set for each model according to Eq.~\eqref{eq:realistic_data}, using new random realization for $\xi$ and $n$. The underlying model for each data set is referred to as the correct model $j$ from now on. The additive noise $n$ is white and Gaussian, i.e $n \hookleftarrow \mathscr{G}\left(n, \sigma_{\text{noise}}\right)$. $\sigma_\text{noise}$ was tuned to a specific signal-to-noise ratio (SNR) which we define as
\begin{equation}
\text{SNR} = \frac{\sigma_{x(t)}^2}{\sigma_{\text{noise}}^2}\,.
\label{eq:SNR}
\end{equation}
Thereby $\sigma_{x(t)}$ refers to the variance of the wave realizations over the displayed period. 
For the observational response $\vec{R}_{\text{obs}}$, see Eq.~\eqref{eq:realistic_data} we assume the unity operator $\vec{R}_{\text{obs}}(t,t')= \delta_{tt'}$. 
\begin{figure}
\centering
	\begin{overpic}[width=.48\textwidth]{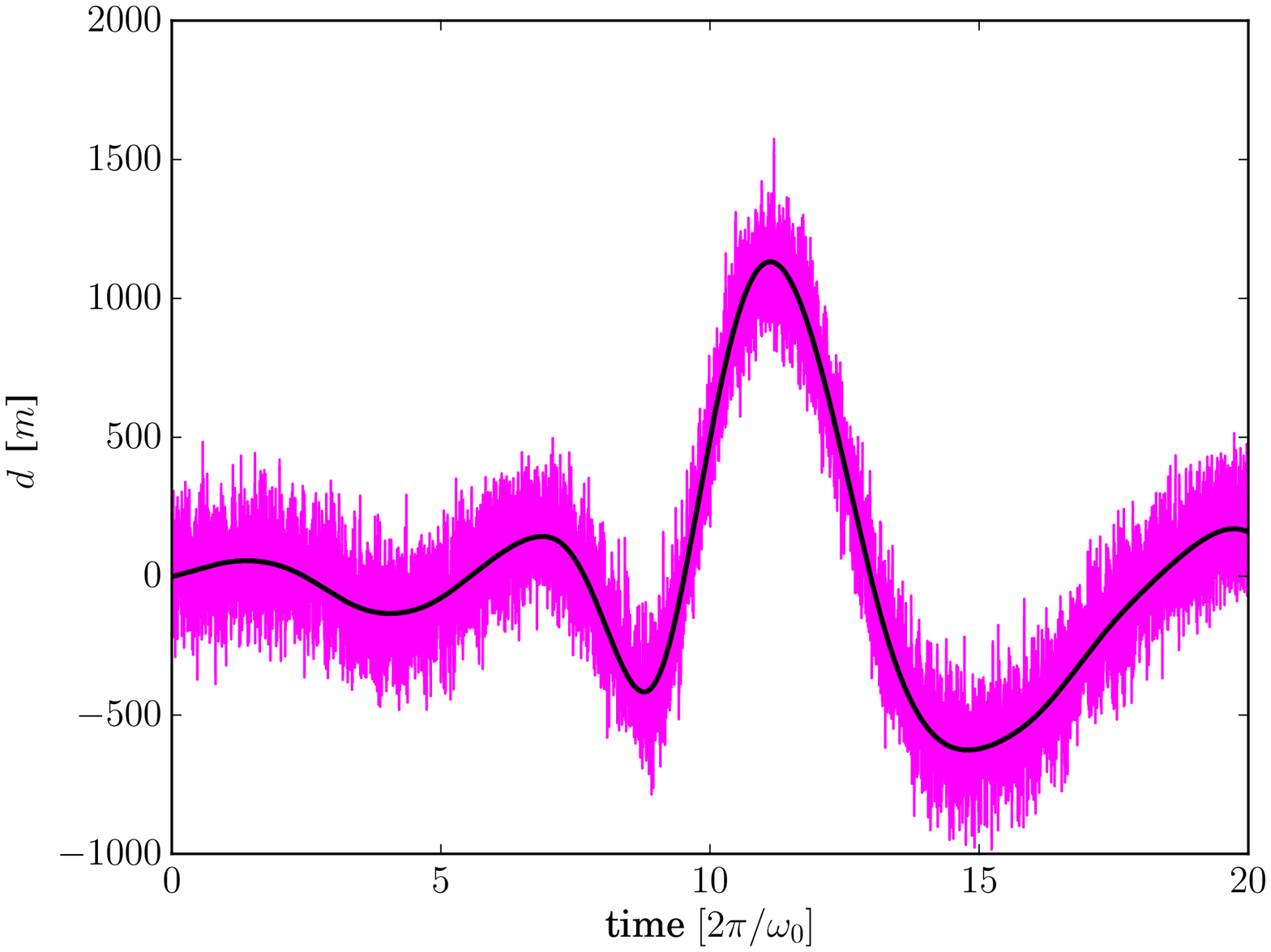} \put(13.5,77){(a)} \end{overpic} \\ 
	\vspace{.5cm}
	\begin{overpic}[width=.48\textwidth]{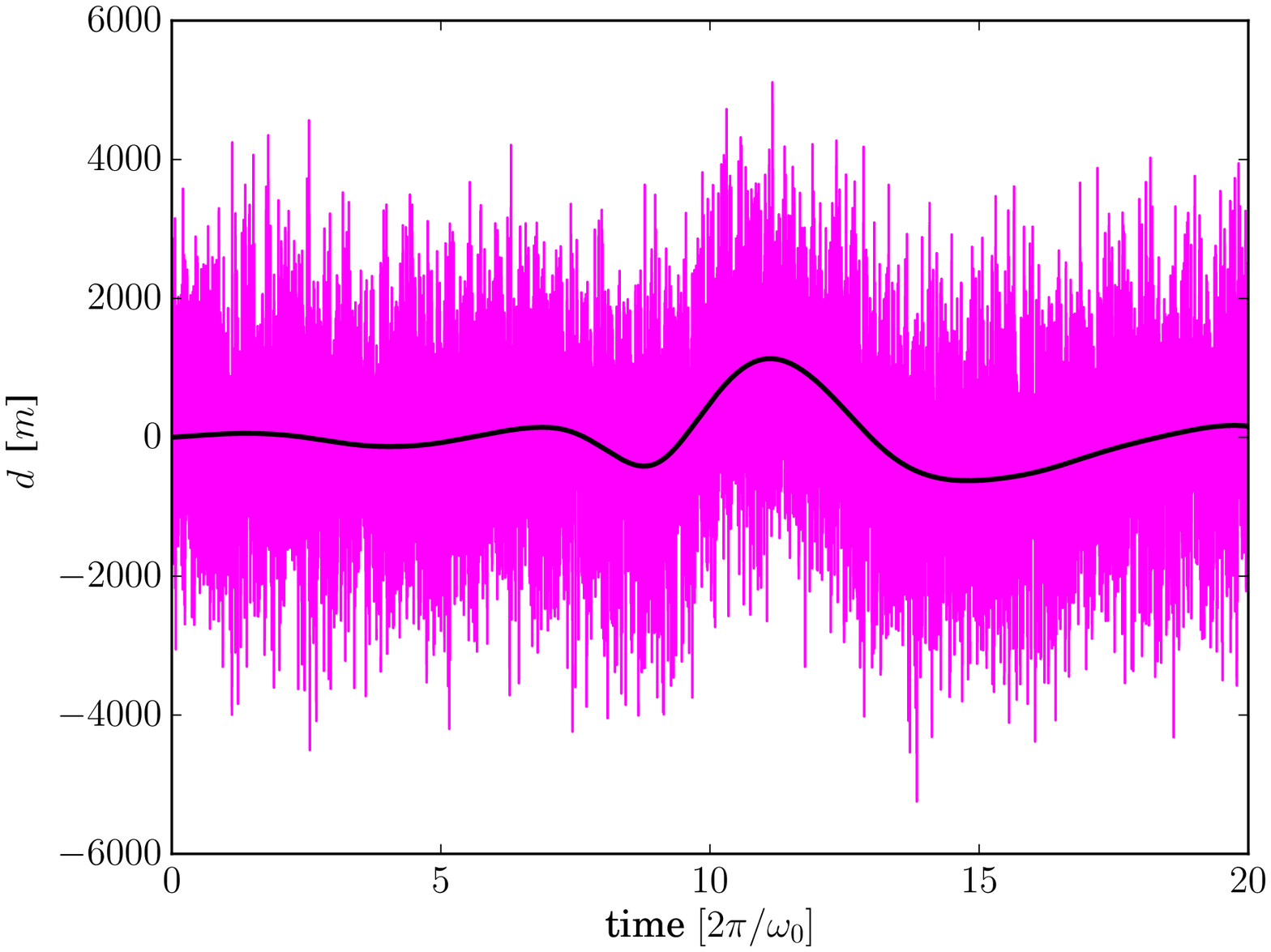} \put(13.5,77){(b)} \end{overpic} \\
	\vspace{.5cm}
	\begin{overpic}[width=.48\textwidth]{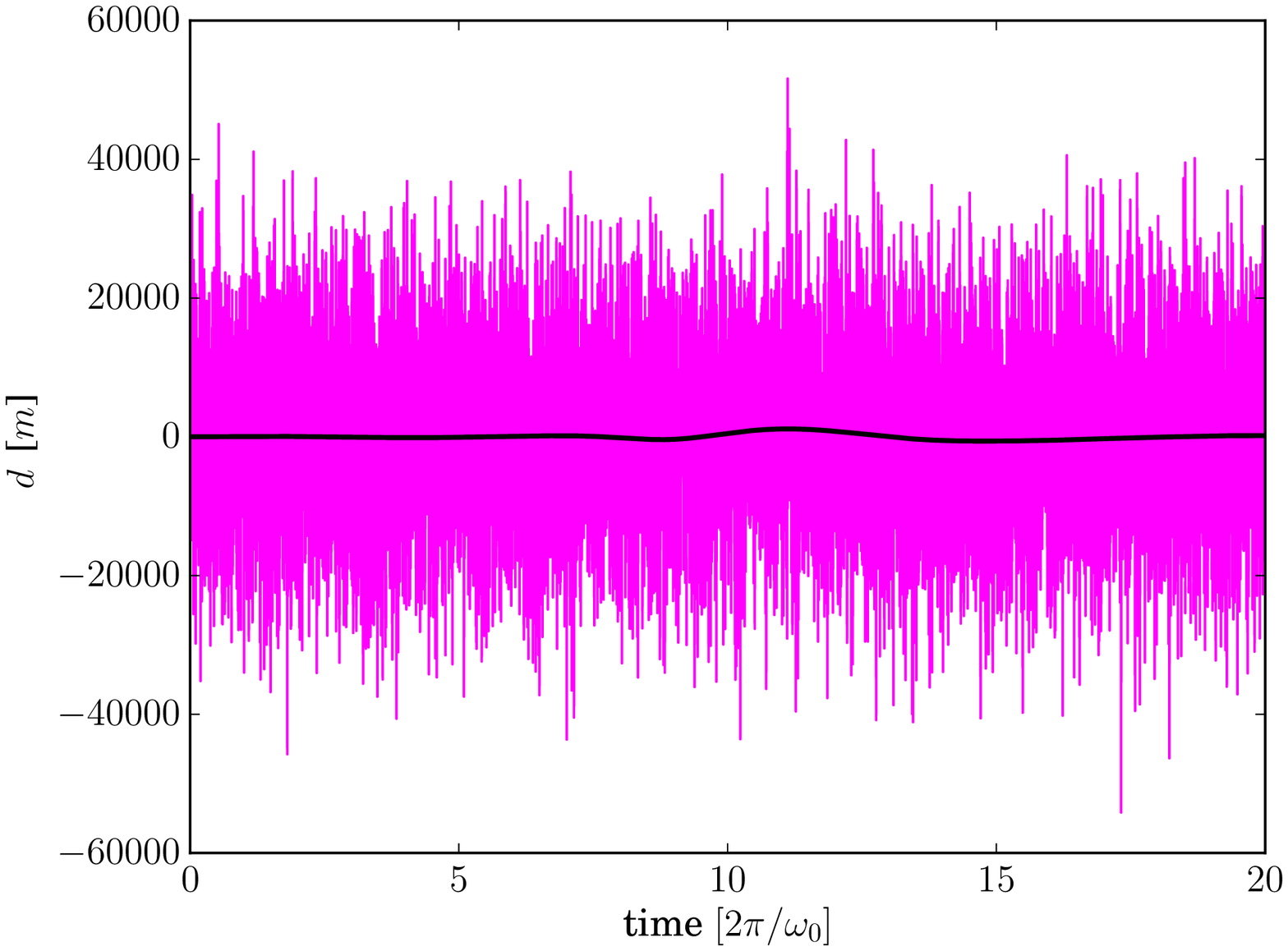} \put(13.5,77){(c)} \end{overpic} \\
	\flushleft 
	\caption{Panel (a), (b), and (c) show data realizations, Eq.~\eqref{eq:realistic_data} of $s_1$. The SNR was tuned to $10$, $0.1$ and $0.001$, respectively.}
\label{fig:data_noise}
\end{figure} 
Fig.~\ref{fig:data_noise} shows three different SNR scenarios for one our waves generated from $s_1$. 
To demonstrate the performance of the model selection algorithm we choose $P\left( s_i \vert \mathscr{S}\right)= \frac{1}{3}$ for all $s_i$, i.e. we did not prefer any model. This allows us to only look at the likelihood of each model instead of its posterior.

In Table \ref{table:model_selection} we give the differences of the log-likelihood 
\begin{align}
\Delta_{ij} &= \text{surprise of model }i - \text{surprise of correct model }j \notag  \\
		&= \mathscr{H}(d\vert s_i) - \mathscr{H}(d\vert s_j) \notag \\
		&= -\log \mathcal{P} (d \vert s_i) + \log \mathcal{P}(d\vert s_j)
\label{eq:delta}
\end{align}
with $i,j \in \{1,2,3\}$, for various SNR. Note that the information Hamiltonian, the negative $\log$ probability, can also be considered as the amount of surprise. The larger the value $\Delta_{ij}$ is, the less plausible an assumed class $i$ is compared to the correct class $j$.
It would be as more surprising from the perspective of the data that the assumed model $i$ is correct compared to the correct $j$. Up to a SNR$=0.01$ all $\Delta_{ij}\geq 0$ and all $\Delta_{ii}=0$, which means that all datasets are correctly classified. If the SNR is worse than $0.01$ the algorithm starts to give misleading classification results, however, only on the $1\sigma$ level, and therefore not with convincing significance.    
\begin{table}
\begin{ruledtabular}
		\begin{tabular}{|l | ccc|}
		SNR=0.001 & $s_1$ assumed & $s_2$ assumed & $s_3$ assumed \\
		\hline
		$s_1$ correct& 0 &  5.10 & -0.82 \\
		$s_2$ correct& 7.17& 0&  7.62 \\
		$s_3$ correct& 1.10& 0.65 & 0\\
		\hline
		SNR=0.01 &&&\\
		\hline
		$s_1$ correct& 0 &  45.6 & 5.14 \\
		$s_2$ correct& 39.4& 0&  39.6 \\
		$s_3$ correct& 6.02& 36.5 & 0\\
		\hline
		SNR=0.1&&&\\
		\hline
		$s_1$ correct& 0 &  374 & 66.4 \\
		$s_2$ correct& 395& 0&  353 \\
		$s_3$ correct& 60.4& 209 & 0\\
		\hline
		SNR=10&  & &  \\
		\hline
		$s_1$ correct& 0&36100& 11200 \\
		$s_2$ correct& 10500& 0&8720 \\
		$s_3$ correct&311&631&0\\
		\hline
		\end{tabular}
	\end{ruledtabular}
\caption{The four tables above show the performance of the model selection algorithm, for different the signal-to-noise ratio (SNR) of the analyzed data. The printed values denote the relative differences of the likelihood according to Eq.~\eqref{eq:delta}. }
\label{table:model_selection}
\end{table}
As one intuitively expects the algorithm performs better in case of high SNR, because the absolute differences between the likelihoods are the largest within this regime. 

\section{Conclusion}
\label{sec:conclusion}
We have established the dynamic system classifier (DSC) algorithm for model selection between dynamic systems. The algorithm consists of two steps. First, it analyzes training data from system classes to construct abstract classifying information for each model class. For distinguishing oscillating systems, a natural basis is the systems' time dependent frequency and damping factor evolution. 
In the second step the algorithm confronts data with the previously learned models and states the probability, which of the learned models $s$ explains the data best. With these capabilities, DSC is a powerful tool to analyze stochastic and dynamically evolving systems. It can abstract a set of sample timelines into characteristic coefficients which encode a non-stationary correlation structure of the signals of the class. 

The theoretical foundations of the first step of DSC is based on a hierarchical Bayesian parameter model within the framework of information field theory.  The model needs \textit{a priori} very few assumptions that account for the statistics and correlations of the two components, $\beta_t$ and $\gamma_t$. Both of them are assumed to obey multivariate Gaussian statistics, whose temporal covariance is described by a power spectrum. The power spectra of both parameter fields are expected to be unknown \textit{a priori}. Therefore, they are also reconstructed from the data, by using the critical filter \cite{2011PhRvD..83j5014E}. This approach is based on the introduction of hyper priors as well as a spectral smoothness enforcing prior \cite{2013PhRvE..87c2136O}. The strength of our proposed and tested DSC algorithm is that it only depends on very few parameters, which can all be motivated \textit{a priori}, and all of them are equally important for the inference.  

The classification ability of the DSC-algorithm has successfully been demonstrated in realistic numerical tests, which showed that one needs at least three data realization of each system class in order to be able to sufficiently characterize the frequency and the damping factor evolution. This is due to the high degeneracy of our problem, as we are trying to reconstruct two parameters and their correlation structures from a few timelines. After learning a number of system classes in terms of their characteristic frequency and damping factor evolvements, the algorithm properly classified realistic measurement data. Down to a SNR $=0.01$ the algorithm determined for all three test models the correct underlying system class with high significance. 

The DSC-algorithm should be applicable to a wide range of inference problems. Concrete examples may be found within in the field of gravitational wave physics. 
\begin{acknowledgments}
We thank Theo Steininger, Sebastian Dorn, Marco Selig and two anonymous referees for fruitful discussions and beneficial comments. 
We also thank the DFG Forschergruppe 1254 "Magnetisation of Interstellar and Intergalactic Media: The Prospects of Low-Frequency Radio Observations" for an inspiring workshop in Summer 2015. Most of the results in this publication have been calculated using the $\textsc{Nifty}$ package \cite{2013A&amp;A...554A..26S}. 
\end{acknowledgments}

\bibliography{bibliography}

\end{document}